# Advancing Building Energy Modeling with Large Language Models: Exploration and Case Studies

Liang Zhang[1,2], Zhelun Chen[3], Vitaly Ford[4]
1. University of Arizona, Tucson, AZ
2. National Renewable Energy Laboratory, Golden, CO
3. Drexel University, Philadelphia, PA
4. Acadia University, Glenside, PA

## Highlights

- Exploring and Highlighting LLM's transformative impact on BEM applications
- Case studies illustrate LLM's role in enhancing BEM's efficiency and accessibility
- Selecting suitable LLM techniques critical for optimizing BEM process efficiency
- Challenges in LLMs' computational demands and consistency addressed by research

## Abstract

The rapid progression in artificial intelligence has facilitated the emergence of large language models like ChatGPT, offering potential applications extending into specialized engineering modeling, especially physics-based building energy modeling. This paper investigates the innovative integration of large language models with building energy modeling software, focusing specifically on the fusion of ChatGPT with EnergyPlus. A literature review is first conducted to reveal a growing trend of incorporating large language models in engineering modeling, albeit limited research on their application in building energy modeling. We underscore the potential of large language models in addressing building energy modeling challenges and outline potential applications including simulation input generation, simulation output analysis and visualization, conducting error analysis, co-simulation, simulation knowledge extraction and training, and simulation optimization. Three case studies reveal the transformative potential of large language models in automating and optimizing building energy modeling tasks, underscoring the pivotal role of artificial intelligence in advancing sustainable building practices and energy efficiency. The case studies demonstrate that selecting the right large language model techniques is essential to enhance performance and reduce engineering efforts. The findings advocate a multidisciplinary approach in future artificial intelligence research, with implications extending beyond building energy modeling to other specialized engineering modeling.

## 1. Introduction

Buildings are significant contributors to global energy consumption and carbon emissions, responsible for approximately 30% of the world's energy use and 26% of CO2 emissions [1]. Buildings represent a critical sector in the global pursuit of decarbonization and reduction of greenhouse gas emissions [2]. Building Energy Modeling (BEM) plays a pivotal role in this endeavor. BEM is a computational technique that uses algorithms to simulate and predict the energy consumption of buildings based on various parameters, such as architectural design, materials, operational schedules, and local climate. It serves as a powerful tool for architects, engineers, and policymakers, aiding in the design and operation of energy-efficient buildings, as well as in the formulation of effective building codes and standards. By optimizing energy use and implementing renewable energy systems, BEM facilitates the path to building decarbonization.

BEM, at its core, is a highly technical and specialized discipline, steeped in a need for extensive knowledge and experience. This necessity stems from the multifaceted and interconnected nature of building science and the diverse range of systems that underpin a building's operations, particularly in the field of heating, ventilation, and air conditioning (HVAC). Users of BEM must understand the fundamentals of these systems, as well as the principles of physics that govern their interactions, in order to accurately capture building characteristics and thus correctly model its energy consumption. This deep understanding needs to be paired with proficiency in specific BEM software. Each of these software packages comes with its own nuances, language, and operational complexities. Mastering these tools demands a significant investment of time and effort, often deterring those who lack the necessary background or resources from effectively leveraging BEM in their work. Furthermore, the sophistication of modern buildings, equipped with complex mechanical systems and novel materials, adds to the challenges faced by BEM practitioners. Buildings are no longer standalone entities but parts of broader energy networks, connected to other buildings and infrastructures. This expanded scope, coupled with an ever-increasing push for sustainability, means that users of BEM now must possess an even more diversified range of expertise, from understanding emerging technologies to interpreting complex regulations and codes. All these factors make BEM an expertise-intensive area, requiring a deep and broad knowledge base that spans multiple disciplines, making it an intricate field to navigate for newcomers and even some experienced professionals.

The rapid progression in the field of artificial intelligence (AI) has facilitated the emergence of Large Language Models (LLMs) like ChatGPT, offering potential applications extending into specialized engineering modeling. Most engineering models are programmable, and the advent of tools like GitHub Copilot [3] empowered by LLMs has dramatically transformed the programming landscape by lowering the barriers to generating programming code for specialized engineering modeling. Power system engineers utilize physics-based modeling and simulation tools, such as PSS/E, PSCAD, PowerWorld, and CyME, which could be called upon by LLMs to solve complex problems [3]. Rios et al. [4] utilize ChatGPT 4 to co-develop a framework to select and train surrogate models for engineering optimization tasks. They interact with ChatGPT to outline a process and software to support the selection and application of regression techniques based on the characteristics of the available data and target application. Patterson [5] reflected on how LLMs can be applied in systems engineering and mechanical design, particularly emphasizing their use in enhancing engineering modeling through pattern recognition and data interaction, which improves the development and refinement of engineering tools and techniques. Although BEM is one particular type of specialized engineering modeling, only a few papers are found to discuss this topic. Jiang et al. [6] successfully translated natural language descriptions of buildings into established models encompassing various geometries, occupancy scenarios, and equipment loads by utilizing a fine-tuned LLM. Despite achieving a perfect accuracy rate of 100%, the study's approach is limited to simplified geometries and building descriptions, necessitating enhancements for more complex modeling scenarios.

The integration of LLMs in BEM holds significant potential due to its transformative impact on human-machine interactions. Traditionally, user engagement with complex machinery or systems, like BEM software, has been constrained by a steep learning curve and the need for specialist knowledge. However, LLMs, with their ability to comprehend and generate reasonable natural language, can significantly streamline these interactions, making them more accessible and intuitive. LLMs essentially serve as an interface, allowing users to communicate with the BEM software using natural language. This drastically lowers the technical barrier, enabling those without specialist knowledge to interact with BEM systems. For instance, a user could instruct the system to modify certain parameters or request an interpretation of the simulation results in simple, everyday language, and the LLM can translate these instructions into actions or provide explanations.

Moreover, LLMs are not limited to merely simplifying interactions; they can also contribute to knowledge enhancement. Given their vast training data encompassing various topics, LLMs can offer valuable insights, explanations, or suggest best practices related to BEM. They can potentially serve as an intelligent assistant, guiding users through complex BEM tasks, enhancing their understanding, and helping them make informed decisions. This shift in how users interact with BEM tools could democratize access to these systems, broadening their application and thus contributing more effectively to decarbonization goals.

To enhance the understanding of the combination and application of LLM and BEM in this study, we provide a detailed research roadmap. Initially, we conducted a literature review to identify current trends and gaps in LLM applications in BEM. We then identified major challenges in BEM, such as the complexity of input data generation and the difficulty in interpreting simulation outputs. To address these issues, we selected appropriate LLM techniques, including prompt engineering, multi-agent systems, and retrieval-augmented generation (RAG). We designed and implemented three case studies focusing on simulation input generation, simulation output visualization, and knowledge extraction and training. These studies demonstrated the potential of LLMs to automate and optimize BEM tasks. The results highlighted the effectiveness and efficiency of LLM-based workflows compared to traditional human modeling efforts. Finally, we discussed the limitations and proposed future research directions to further enhance the integration of LLMs in BEM.

The paper is structured as follows. In Section 2, we review the development and characteristics of LLM and its application in facilitating the usage of specialized and professional software; then we summarize the promising applications of LLMs in BEM. In Section 3, we design preliminary case studies to demonstrate the effectiveness of potential LLM applications in BEM. In Section 4 and 5, we discuss the results of the case studies and conclude with an outlook on future trends and developments in the application of LLMs in BEM.

## 2. LLM in Engineering Modeling and BEM

At the point of writing the paper, we found few paper discussing the topic of LLM in BEM. To understand the existing work that can benefit this particular topic, we review the papers from a broader perspective. Since BEM is a type of interaction between humans and software requiring expert knowledge, it is worth investigating how LLM has already helped with benefiting the usage of specialized and professional software-based modeling requiring expert knowledge (Section 2.1); then we summarize the applications of LLM in BEM based on the advantages of LLMs and the pressing needs in the BEM field from authors' domain knowledge (Section 2.2).

### 2.1 Applications of LLM in General Engineering Modeling with BEM Context

#### 2.1.1 User Interfaces

LLMs can provide a conversational interface to interact with complex software, simplifying the user experience. For instance, they can enable users to perform tasks using natural language commands, rather than having to navigate complicated menus or learn specific programming languages. Wen et al. [7] use LLM to enable voice-based hands-free user interaction with smartphone apps.

In BEM, a user interface is crucial for ease of use. Consider EnergyPlus [8], where all the user interfaces (e.g., OpenStudio, DesignBuilder, and Ladybug Tools) are graphical user interfaces. However, the forms of interface should not be limited. A well-designed interface empowers BEM users to easily express their modeling needs and receive simulation results in a manner that is most direct and comprehensible, making the natural-language-based user interface a promising field. Currently, there is a lack of such interfaces in BEM, and we posit that their integration into software tools would substantially enhance workflow efficiency.

#### 2.1.2 Code Generation

LLMs have the capability to comprehend programming languages and generate code snippets from natural language prompts. GitHub Copilot (https://github.com/features/copilot), a collaborative effort between GitHub and OpenAI, serves as an AI pair programmer, offering code suggestions while developers write, thus accelerating the coding process, reducing error potential, and offering a learning resource.

LLMs can further automate data and modeling workflow through code generation capabilities. This allows LLM to automate complex processes by coordinating tasks across various software tools. They can interpret instructions given in natural language, create the required commands or scripts, and then perform or arrange the tasks as needed. Many LLM-based tools (e.g., AutoGen [9] and MetaGPT [10]) have already proven their ability in LLM-based workflow automation. In data-intensive fields, LLMs can automate tasks such as data cleaning, preprocessing, analysis, and visualization. They can understand high-level descriptions of the desired data transformations or analyses, generate the necessary code, and provide the results in a user-friendly format.

The automation of simulation tasks is a very important branch of BEM. Currently, the most widely used BEM automation methodology is OpenStudio Measure [11]. OpenStudio Measures are Ruby scripts that extend OpenStudio's functionality, enabling users to customize energy models, implement energy-saving strategies, and automate tasks in a collaborative platform. However, the development of OpenStudio Measure has a very high requirement for the skills of Ruby programming language, OpenStudio, and EnergyPlus, as well as knowledge in building science and building equipment. LLM has great potential to further "automate the automation" by auto-generating Ruby scripts. It will bring the automation of BEM to the next level from data collection, model generation, and simulation results reporting.

The LLM has proven to be highly effective in workflow automation, seamlessly orchestrating tasks across a diverse range of software tools. This capability makes it an ideal candidate for BEM co-simulation tasks. The automation of co-simulation tasks is an integral aspect of code generation, as these tasks are programmed in various software tools, each requiring adherence to specific syntax and formats. BEM co-simulations require the integration of multiple software tools and models to meticulously simulate and analyze a building's energy performance, considering a multitude of factors including HVAC systems, weather conditions, occupant behavior, and the characteristics of the building envelope. The expertise of the LLM in code generation, workflow automation, and data processing positions it as a valuable asset in streamlining and enhancing the efficiency of BEM automation processes.

#### 2.1.3 Documentations, Tutorials, and Training

Documentations, tutorials, and training play a crucial role in the effective and efficient use of any professional software. They serve as the first point of contact for new users and a reference guide for experienced ones. In the past, these resources were static and sometimes difficult to comprehend, especially for complex software. However, the advent of LLM is ushering in a new era of intelligent, dynamic, and interactive user assistance. One of the most exciting capabilities of LLMs is their ability to generate and reorganize content in a way that makes it more accessible and user-friendly. LLMs can produce well-structured documentation, interactive tutorials, and step-by-step guides in real-time, tailored to the specific needs of the user. For instance, an LLM could produce a beginner's guide to complex data analysis software by generating explanations and examples in plain language, or generate a more advanced tutorial focusing on a particular feature or use case based on the user's specific query.

In addition, LLMs also offer real-time support by answering specific questions about software features. Rather than having to sift through a FAQ page or search for a relevant tutorial or question-and-answer forum, users can simply ask the LLM their questions in natural language. The LLM can understand the query, find the most relevant information, and generate a helpful response. This kind of interactive, on-demand assistance can significantly reduce the learning curve associated with complex software, making it more accessible to a broader range of users.

MacNeil et al. [12] reported on their experiences generating multiple code explanation types using LLMs and integrating them into an interactive e-book on web software development. Three different types of explanations – a line-by-line explanation, a list of important concepts, and a high-level summary of the code – were created. Their results show that all explanation types were viewed by students and that the majority of students perceived the code explanations as helpful to them. Su et al. [13] explores the question of how to make software documentation more useful with an LLM. They investigate a general, one-model-fit-all solution through a state-of-the-art LLM (ChatGPT). The paper covers three representative tasks: extracting locking rules from comments, synthesizing exception predicates from comments, and identifying performance-related configurations; it also reveals challenges and opportunities in applying LLMs to system maintenance.

*2.1.4 Error Identification and Troubleshooting*

Error identification and troubleshooting have traditionally been complex processes, requiring specialized knowledge and experience. However, the incorporation of LLM into these systems is transforming how these tasks are performed, making them more efficient and accessible to a broader range of users. LLMs can assist in identifying and troubleshooting errors by interpreting descriptions of issues provided by the users. This involves natural language processing capabilities that allow the AI to understand the user's language, including technical terms and even colloquial or less precise descriptions of problems. The LLMs can then match these descriptions with known errors or issues, helping to pinpoint what may be going wrong. One of the main benefits of using LLMs in error identification is that they can significantly reduce the time taken to understand and diagnose the problem. For example, if a user encounters a software crash, they could describe the issue to the LLM, which would then process this description, correlate it with known bugs or issues, and suggest possible causes for the crash. In terms of troubleshooting, LLMs can provide step-by-step guidance to resolve the identified issues. Based on the identified error, the LLM can generate a list of potential solutions, ordered by their likelihood of success or ease of implementation. This could range from simple solutions like restarting the software to more complex procedures such as modifying specific settings or running certain commands. In each case, the LLM can provide clear, easy-to-follow instructions, making it easier for non-expert users to resolve issues on their own.

Moreover, LLMs can learn from each interaction, thereby enhancing their ability to handle similar issues in the future. This capability allows them to become more effective over time, ultimately improving the efficiency of the troubleshooting process. This debugging process can also be automated and integrated within the software's operational cycle, allowing the system to self-correct iteratively until it operates without faults, thus streamlining the modeling process and enhancing system reliability. Overall, the use of LLMs in error identification and troubleshooting represents a significant leap forward. By enabling rapid diagnosis and resolution of software issues, they not only enhance the user experience but also increase the overall efficiency and reliability of software systems.

Most commercial LLM tools are available for general error identification and troubleshooting. For instance, ChatGPT can assist with debugging by pinpointing and clarifying common errors like syntax or logical mistakes. Unfortunately, similar tools specifically designed for professional software are currently lacking.

In the context of BEM, error identification and troubleshooting have traditionally been complex processes, especially for expert-knowledge-dependent software such as EnergyPlus. Users often have to sift through dense technical documentation or rely on trial-and-error methods to identify and rectify issues, which can be time-consuming and inefficient. However, with the introduction of LLMs, these processes could be significantly streamlined and enhanced.

## 2.2 Promising Applications of LLM in BEM

In this sub-section, we further summarize the advances and advantages of LLMs in the context of the key challenges in BEM, especially its heavy dependency on expert knowledge. We explore and propose several application types of LLMs with case studies to enhance and streamline the BEM process.

### *2.2.1 Simulation Input Generation*

Defining simulation input is a foundational step in BEM, where detailed parameters such as building geometry, material properties, HVAC system configurations, occupancy patterns, and local climate data are defined to represent a building's characteristics for energy modeling. LLMs, equipped with vast knowledge bases and adept natural language processing capabilities, are uniquely positioned to streamline this intricate process. For instance, a user might describe a building's façade as "mostly glass with southern exposure." An LLM, through prompt engineering, can interpret this to generate specific parameters like window-to-wall ratio, glazing type, and solar heat gain coefficients. An LLM can then adeptly transform these descriptions into a structured input format, meticulously populating a BEM input file, such as the Input Data Dictionary (IDD) for EnergyPlus, ensuring all parameters align with the template's requirements. In summary, LLMs significantly enhance the efficiency of setting up BEM by translating natural descriptions into precise simulation inputs, ensuring accurate and streamlined energy analysis.

### *2.2.2 Simulation Output Analysis and Visualization*

BEM has a structured simulation output format, which is very suitable to be processed by LLM with its ability of code generation to automatically conduct data analysis, modeling, and visualization. Besides, the unique challenges of BEM outputs demand more specialized solutions. LLMs, equipped with capabilities of context-aware data interpretation, can not only contribute to data automation but also assist in offering deeper insights and extracting meaningful knowledge from vast simulation datasets. For instance, when analyzing a spike in energy consumption, an LLM might correlate it with specific HVAC activities during peak occupancy hours, offering a nuanced understanding.

### *2.2.3 Conducting Error Analysis*

As simulations grow in complexity, the potential for errors increases, and these errors can manifest in various ways. Some errors, due to violations of basic model assumptions or misconfigurations, can cause the simulation to fail outright. For instance, specifying an impossible combination of materials or an HVAC system operating outside its feasible range might halt an EnergyPlus simulation before it even begins. On the other hand, subtler errors might not stop the simulation but can lead to anomalous results. An incorrectly defined occupancy schedule or a misconfigured shading device might not prevent the simulation from running but could result in unexpected energy consumption spikes or temperature fluctuations. The challenge arises from error complexity and a lack of feedback mechanisms.

LLMs can assist in pinpointing and elucidating these errors. For a complete simulation failure, an LLM might trace the issue to a specific input violation. For anomalous results, it might highlight potential inconsistencies or misconfigurations that led to the unexpected behavior.

While LLMs can identify and explain many known errors, novel or unprecedented issues might be harder to diagnose. The vast array of potential BEM errors, each with its unique characteristics, makes error analysis in tools like EnergyPlus a nuanced task. Continuous fine-tuning of the LLM on the latest BEM datasets and updates is essential. For instance, EnergyPlus has a rich ecosystem of resources like the Engineering Reference, the Input Output Reference, and community forums. An LLM can be trained on these resources to enhance its diagnostic capabilities. When a user encounters an error, the LLM can cross-reference the user's description with known issues from these resources, provide relevant excerpts from user documents, or even suggest similar cases discussed in community forums. Integrating user feedback loops enables LLM to learn from its misses, refining its diagnostic capabilities over time.

### *2.2.4 Co-Simulation*

EnergyPlus is good at the simulation of whole-building energy consumption without detailed and dynamic energy system modeling. Co-simulation in BEM involves the concurrent use of multiple simulation tools, each specialized in a particular domain inside or even outside of a building, to provide more detailed and realistic modeling, for instance, dynamic HVAC system and control modeling using Modelica/MATLAB/Python, detailed occupancy modeling tool from studies such as [14], weather forecasting modules [15], dynamic insulation such as phase change materials [16], distributed energy systems [17], and energy storage system modeling [18].

The integration of LLMs in co-simulation processes can streamline the coordination between these tools. LLMs can potentially understand the intricacies of each tool and ensure that data is seamlessly transferred and interpreted across platforms. However, challenges arise in co-simulation. Ensuring real-time synchronization between different tools, managing data consistency, and handling potential conflicts in overlapping domains are all intricate tasks. Additionally, the sheer diversity of tools, each with its own set of assumptions, parameters, and output formats, can complicate the integration process.

*2.2.5 Simulation Knowledge Extraction and Training*

Efficient and comprehensive documentation and training pose a significant challenge in BEM. Consider EnergyPlus as an instance; substantial efforts have been invested by federal agencies, professional organizations, and companies to create helpful resources. However, most training and tutorials of EnergyPlus are limited to the form of 1) static and web-based documentation, 2) online and offline training sessions, 3) question-and-answer sites, and 4) online encyclopedias. Since LLMs are revolutionizing how we understand and interact with the documentation and the tutorials of expert software through their ability to generate, reorganize, and present information in an intelligent and user-friendly manner, they are not only simplifying the use of complex software but also enhancing the learning experience for users of all levels. The result is to provide a more inclusive, efficient, and effective learning and documenting experience for BEM. Besides, through the fast-developing BEM technologies, the topic of up-to-date knowledge is extremely relevant and important. LLMs can stay updated with new knowledge, so they can always provide accurate information and support, something that static documentation can struggle with.

*2.2.6 Simulation Optimization*

Optimizing a building's energy performance is a multifaceted endeavor, drawing heavily on the processes detailed in earlier sections. At its core, optimization refines the myriad parameters that define a building's energy model to achieve the best possible outcomes.

For instance, while Section 2.2.1 discussed how an LLM can assist users in defining parameters based on their descriptions, in the context of optimization, the LLM's role shifts slightly. Using the building's façade example, instead of merely interpreting a user's description, the LLM might proactively suggest specific parameters, such as window-to-wall ratio, glazing type, or solar heat gain coefficients, to optimize. These suggestions would be informed by a combination of factors: extensive datasets of similar building configurations and their performance metrics, best practices in architectural and engineering design, historical trends in energy consumption, predictive models of future energy needs, and even feedback loops from real-world building performance post-occupancy.

Ultimately, the goal of simulation optimization is to harmonize energy efficiency with building functionality and occupant comfort or well-being. While LLMs can provide invaluable data-driven insights and suggestions, the intricate nuances of building design, occupant behaviors, and real-world conditions underscore the irreplaceable value of human judgment in the decision-making process.

As we transition into the case studies, it is essential to emphasize that LLMs are tools designed to augment our expertise, not replace it. Their role is to assist and enhance, while the final decisions and creative insights remain inherently human.

## 3. Case Studies

In this section, we design three case studies to demonstrate the effectiveness of potential LLM applications in BEM. In conducting our case studies, we employ three key methodologies harnessing the capabilities of LLMs: 1) prompt engineering, 2) multi-agent LLMs, and 3) retrieval-augmented generation (RAG). The prompt engineering method revolves around carefully crafting prompts or instructions to guide the LLM in executing desired tasks. This method capitalizes on the LLM's ability to interpret and respond to natural language prompts without requiring specific model alterations. It involves a deep understanding of how the model processes and responds to different types of prompts, and leveraging this understanding to generate accurate and effective outcomes. On the other hand, multi-agent LLMs incorporate multiple LLMs working collaboratively to solve complex problems or perform intricate tasks. This approach capitalizes on the collective intelligence and diverse capabilities of multiple LLMs, allowing for more comprehensive and nuanced problem-solving. Both these methodologies offer unique advantages and can be leveraged according to the specific requirements of the task at hand. While the prompt Engineering method can be utilized quickly and efficiently, the multi-agent LLMs offer superior performance for tasks that demand a combination of specialized knowledge, creativity, and collaborative decision-making, providing a robust solution that often surpasses the capabilities of a single LLM. Lastly, RAG uses the model's advanced natural language processing capabilities to perform in-depth searches, extracting contextually relevant information from vast datasets. This approach is crucial for BEM tasks that require a comprehensive understanding of complex subject matter and the synthesis of data from multiple sources to produce informed and precise conclusions.

## 3.1 Simulation Input Generation

In Section 3.1, we will apply LLM and its relevant techniques to generate and modify Input Data File (IDF) objects and files as the input of EnergyPlus. In Section 3.1, we use ChatGPT-4 July 6, 2023, Version.

### 3.1.1 Single Object Generation

We first use LLM to generate a people object by telling the LLM that "Generate a 'People' object for me. I want it to be defined by 'Number of People' which is set to 10, and set other field values either default or blank." We first directly send this request to LLM without prompt engineering, and the output is shown as follows.

```
People,
  People_1, !- Name
  , !- Zone or ZoneList Name
  , !- Number of People Schedule Name
  , !- Number of People Calculation Method
  10, !- Number of People
  , !- People per Zone Floor Area (person/m2)
  , !- Zone Floor Area per Person (m2/person)
  0.3, !- Fraction Radiant
  AUTOCALCULATE, !- Sensible Heat Fraction
  Activity Level Schedule Name, !- Activity Level Schedule Name
  3.82E-8, !- Carbon Dioxide Generation Rate (m3/s-W)
  Yes, !- Enable ASHRAE 55 Comfort Warnings
  ZoneAveraged, !- Mean Radiant Temperature Calculation Type
  , !- Surface Name/Angle Factor List Name
  Work efficiency schedule name, !- Work Efficiency Schedule Name
  Clothing insulation schedule name, !- Clothing Insulation Calculation Method Schedule Name
  , !- Clothing Insulation Schedule Name
  , !- Air Velocity Schedule Name
  , !- Thermal Comfort Model 1 Type
  , !- Thermal Comfort Model 2 Type
  , !- Thermal Comfort Model 3 Type
  , !- Thermal Comfort Model 4 Type
  , !- Thermal Comfort Model 5 Type
  , !- Thermal Comfort Model 6 Type
  , !- Thermal Comfort Model 7 Type
```

The generated people object is partially correct. It can be seen that the key field "Number of People Calculation Method" is left blank, meaning that LLM does not capture the people calculation method from the prompt. Besides, the value of "Enable ASHRAE 55 Comfort Warnings" is "yes" instead of "no" (default value), which is against the requirement in the prompt.

We try to use prompt engineering to improve the accuracy of the object generated. We designed a prompt engineering script shown below.

```
Set the temperature to 0
The user wants to automatically generate text-based idf objects for EnergyPlus Simulation. The
format of a certain type of object called "People" is provided in triple back-ticks.
Here is the users' prompt below:
"Generate a "People" object for me. I want the people object to be defined by "Number of People"
which is set to 10, and make other fields value to be either default or blank. "
Some general rules when generating the object:
- If the field is not mandatory and is not defined by the user, please do not put a value in the
field.
- If the field is mandatory but not defined by the user, please set it to "TBD" + a 10-digit random
number
- Even though some fields are optional or not defined by the user, keep them in the object
definition.
- In the object generated, comment after "!" for each field about why the value is set or is left
blank
- If the rules above conflict with the users' prompt, prioritize the rules here and inform the user
there is a conflict
In the end after the object is generated, tell the user:
- Which fields are set by user
```

```
- Which fields are mandatory but not defined
- Which fields are set to their default values
- Which fields are left blank
```
People,
\memo Sets internal gains and contaminant rates for occupants in the zone.
\memo If a ZoneList, SpaceList, or a Zone comprised of more than one Space is specified
\memo then this definition applies to all applicable spaces, and each instance will
\memo be named with the Space Name plus this Object Name.
\min-fields 10
! \field Name
…(full IDD file omitted because of the word limit)
```
```

*In the first paragraph*, we define "temperature" as a parameter that controls the randomness of LLM's output, with a range from 0 to 1. A lower temperature results in more deterministic responses, essential for rule-based tasks such as EnergyPlus object creation. Therefore, we set the temperature to 0. *In the second paragraph*, we provide ground truth to the task by referring to the IDD file. EnergyPlus objects are defined by the IDD file, which provides the structure and format of input data required by the simulation program. *The third paragraph* provide a placeholder for the user's request, in this case, to generate a people object with 10 people. *The fourth paragraph* defines the rules for object generation to (1) guarantee the object aligns precisely with user-defined information and (2) prevent syntax errors by ensuring all obligatory fields are accounted for. Furthermore, we instruct the LLM to annotate field values with comments elucidating the rationale behind its decisions, thereby providing modelers with a transparent view of the LLM's decision-making process.

The output with prompt engineering is shown below. The output format is correct and the values in all fields follow the rules we defined in the prompt template.

```
People,
  TBD1234567890, !- Name - Mandatory but not defined by user, so set to "TBD"
  TBD2345678901, !- Zone or ZoneList or Space or SpaceList Name - Mandatory but not defined
  TBD3456789012, !- Number of People Schedule Name - Mandatory but not defined
  People, !- Number of People Calculation Method - Default value used as it's not defined by user
  10, !- Number of People - Defined by user
  , !- People per Floor Area - Not mandatory and not defined by user, so left blank
  , !- Floor Area per Person - Not mandatory and not defined by user, so left blank
  0.3, !- Fraction Radiant - Default value used as it's not defined by user
  AUTOCALCULATE, !- Sensible Heat Fraction - Default value used as it's not defined by user
  TBD4567890123, !- Activity Level Schedule Name - Mandatory but not defined
  3.82E-8, !- Carbon Dioxide Generation Rate - Default value used as it's not defined by user
  No, !- Enable ASHRAE 55 Comfort Warnings - Default value used as it's not defined by user
  ZoneAveraged, !- Mean Radiant Temperature Calculation Type - Default value used as it's not defined by user
  , !- Surface Name/Angle Factor List Name - Not mandatory and not defined by user
  , !- Work Efficiency Schedule Name - Not mandatory and not defined by user
  ClothingInsulationSchedule, !- Clothing Insulation Calculation Method - Default value used as it's not defined by user
  , !- Clothing Insulation Schedule Name - Not mandatory and not defined by user
  , !- Air Velocity Schedule Name - Not mandatory and not defined by user, so left blank
  , !- Thermal Comfort Model 1 Type - Not mandatory and not defined by user, so left blank
  , !- Thermal Comfort Model 2 Type - Not mandatory and not defined by user, so left blank
  , !- Thermal Comfort Model 3 Type - Not mandatory and not defined by user, so left blank
  , !- Thermal Comfort Model 4 Type - Not mandatory and not defined by user, so left blank
  , !- Thermal Comfort Model 5 Type - Not mandatory and not defined by user, so left blank
  , !- Thermal Comfort Model 6 Type - Not mandatory and not defined by user, so left blank
  , !- Thermal Comfort Model 7 Type - Not mandatory and not defined by user, so left blank
  , !- Ankle Level Air Velocity Schedule Name - Not mandatory and not defined by user, blank
  15.6, !- Cold Stress Temperature Threshold - Default value used as it's not defined by user
  30, !- Heat Stress Temperature Threshold - Default value used as it's not defined by user
```

Besides, the reasoning behind the value is also correctly explained in the comments and in the generated explanation texts. The accuracy of the output is obviously improved compared with that without prompt engineering. After filling TBD values with actual values, it can be run in EnergyPlus without any error.

```
Here's a summary:
Fields set by user: "Number of People"
Mandatory fields not defined: "Name", "Zone or ZoneList or Space or SpaceList Name", "Number of
People Schedule Name", "Activity Level Schedule Name"
```

```
Fields set to default values: "Number of People Calculation Method", "Fraction Radiant", "Sensible
Heat Fraction", "Carbon Dioxide Generation Rate", "Enable ASHRAE 55 Comfort Warnings", "Mean Radiant
Temperature Calculation Type", "Clothing Insulation Calculation Method", "Cold Stress Temperature
Threshold", "Heat Stress Temperature Threshold"
Fields left blank: "People per Floor Area", "Floor Area per Person", "Surface Name/Angle Factor
List Name", "Work Efficiency Schedule Name", "Clothing Insulation Calculation Method Schedule
Name", "Clothing Insulation Schedule Name", "Air Velocity Schedule Name", "Thermal Comfort Model 1
Type", "Thermal Comfort Model 2 Type", "Thermal Comfort Model 3 Type", "Thermal Comfort Model 4
Type", "Thermal Comfort Model 5 Type", "Thermal Comfort Model 6 Type", "Thermal Comfort Model 7
Type", "Ankle Level Air Velocity Schedule Name"
```

Although not explored in this paper, we can further add different types of improvements to the prompt, e.g., "assume the role of the best assistant in IDF object generation", "reason step-by-step and logically at all times", "review generated output in terms of errors and fix them", and "iteratively improve output until it is correct and complete." [19]

In summary, we observed the necessity of prompt engineering in creating a specific EnergyPlus object based on natural language input. In terms of the time spent on object creation, the user takes less than a minute to write a prompt in the placeholder of the prompt template we designed in LLM; furthermore, the user does not need expertise in the IDF object.

### 3.1.2 Whole IDF Modification

We will further investigate whether LLM can deal with the complete IDF file in this section. Since the generation of a complete IDF file requires too much information using the natural language input, we just focus on revising the existing IDF based on the user's requirements in the case study.

We will use multi-agent LLM techniques. The diagram is shown in Figure 1. It consists of a central LLM agent and several LLM task agents. The Central LLM agent communicates with the user, plans sub-tasks, assigns sub-tasks to specialized LLM task agents, aggregates the results from LLM task agents, and sends results to users.

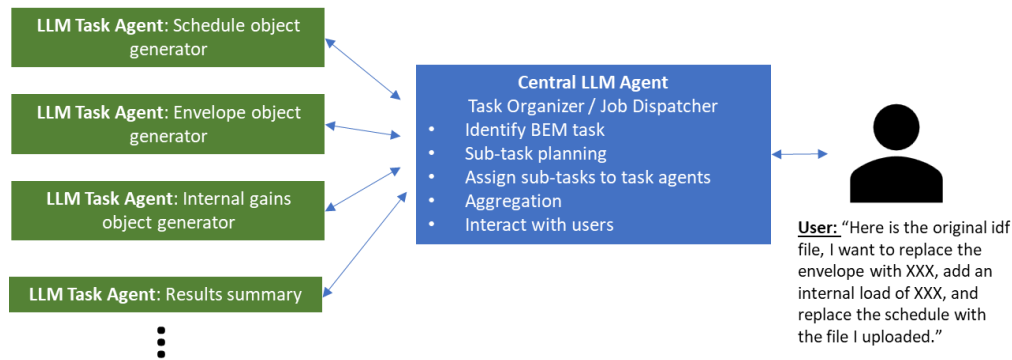

*Figure 1. Multi-agent LLMs for IDF modification*

The Central LLM Agent is based on the GPT-4 Advanced Data Analysis plugin (September 25, 2023, Version), which supports the upload of complete IDF files. The Central LLM Agent is based on the following prompt template.

```
You are the Central LLM Agent in a multi-agent LLMs used to modify idf files based on user's
input. Here are your tasks:
1. Ask for idf files and the modification requirement from user.
2. Based on the user's input, identify the relevant objects and extract them as texts.
3. Send objects in the form of text to the correspondent LLM task agents.
4. Wait for the feedbacks from all agents
5. Aggregate the feedback, correct object reference, and generate a modified idf file and send it
to user.
```

The LLM task agents are in the similar structure with the engineered prompt template for people object creation introduced earlier. The only two differences are 1) the placeholder gets input from the Central LLM Agent, and 2) adding an extra step to return the results to the Central LLM Agent.

All agents are defined, coordinated, and connected through the AutoGen framework [9]. Microsoft's AutoGen is a platform engineered to propel the creation of LLM applications by facilitating multi-agent dialogues. It allows diverse agents to engage in discussions with each other to address tasks, with the agents being adaptable, capable of conversation, and structured to effortlessly accommodate human involvement across different modes utilizing blends of LLMs, human contributions, and tools.

In the case study, we use a simplified IDF file from EnergyPlus official document InterfaceDeveloper.pdf Appendix A. Simple IDF file as a starting point. The user's request is shown as follows.

```
Here is the IDF file and I want to 1) create a people object for the only zone, 2) add a schedule
for exterior lights that starts at 6:00 PM and ends at 6:00 AM, and 3) change the R13 layer to be
R15.
```

Below is the excerpt from the generated IDF file. Due to the page limit, we only show the schedule and construction since the people object is already shown in 3.1.1.

```
Schedule:Compact,
  NewSchedule,          !- Name
  Fraction,             !- Schedule Type Limits
  Through: 12/31,       !- Field 1
  For: AllDays,         !- Field 2
  Until: 06:00,0,       !- Field 3
  Until: 18:00,1,       !- Field 4
  Until: 24:00,0;       !- Field 5
…
Exterior:Lights,
  ExtLights,        !- Name
  NewSchedule,      !- Schedule Name
  5250,             !- Design Level {W}
  AstronomicalClock, !- Control Option
  Grounds Lights;   !- End -Use Subcategory
…
Material:NoMass,
  R13LAYER,         !- Name
  Rough,            !- Roughness
  2.64,             !- ThermalResistance{m2K/W}
  0.9000000,        !- Thermal Absorptance
  0.7500000,        !- Solar Absorptance
  0.7500000;        !- Visible Absorptance
```

In terms of construction material, although it is still called R13Layer, the thermal resistance is updated from 2.29 to 2.64 $m^2K/W$ (R15), which correctly reflect user input. For exterior lighting, the new schedule is correctly created and successfully applied in exterior lights. We observed the effectiveness of multi-agent LLM applications in multiple inputs generation and revisions for BEM, which is a common task for BEM modelers.

*3.1.3 Whole IDF Generation*

In this case study, an LLM planning workflow is designed to handle a more complex BEM task. The goal is to generate an IDF file using the required IDF object information extracted from building descriptions and/or specifications.

As shown in Figure 2, the process starts with two main inputs: 1) a natural-language description of the building, considered as unstructured data, and 2) an initial IDF file containing geometry, zone information, and mandatory IDF objects such as "Version", "SizingPeriod:DesignDay", and "RunPeriod." Geometry generation is excluded from this paper due to its complexity and heavy reliance on graphical user interfaces, which are challenging for LLMs to generate.

The workflow begins by collecting natural-language descriptions of the building from modelers, detailing internal loads, envelope materials, energy systems, and more. Next, the IDF Object Type and Description Extraction Agent (Agent 1) systematically identifies potential IDF objects for materials, loads, and HVAC systems within the description, ensuring no element is overlooked. Then, the Single IDF Object Generator Suite (Agent 2) uses the prompt templates introduced in Section 3.1.1 to systematically generate IDF objects from these descriptions, ensuring all IDF object types are correctly formatted. Finally, the generated objects are merged with the initial IDF file, debugged in Agent 3, and form the final error-free IDF file.

Agent 1 extracts specific types of information, including 1) materials/construction, 2) internal loads, and 3) HVAC systems. Each agent has placeholders for additional information to complete the prompts, such as {user_description} (the building description) and {common_object_list} (a user-defined list suggesting the object type for creation based on domain knowledge, which is optional). After extracting this information, Agent 1 produces a dictionary or list containing the names of the material/construction objects and their text-based descriptions.

```
Agent_1 = f"""I want to process user description of their building to create a dictionary about 1)
material and construction, 2) internal load (people, lights, electric equipment, infiltration),
and 3) HVAC system. The dictionary's keys are the EnergyPlus IDF object type and its name to be
created in tuple format (object type in str, object name in str, object name cannot be the same as
object type), and the values to be detailed description or requirement of the IDF object, in string
format.

Only generate object type mentioned in commonly used internal load objects list attached at the
end of the prompt. Here is the user description: {user_description}

Commonly used objects list: {common_internal_load_object_list}
"""
```

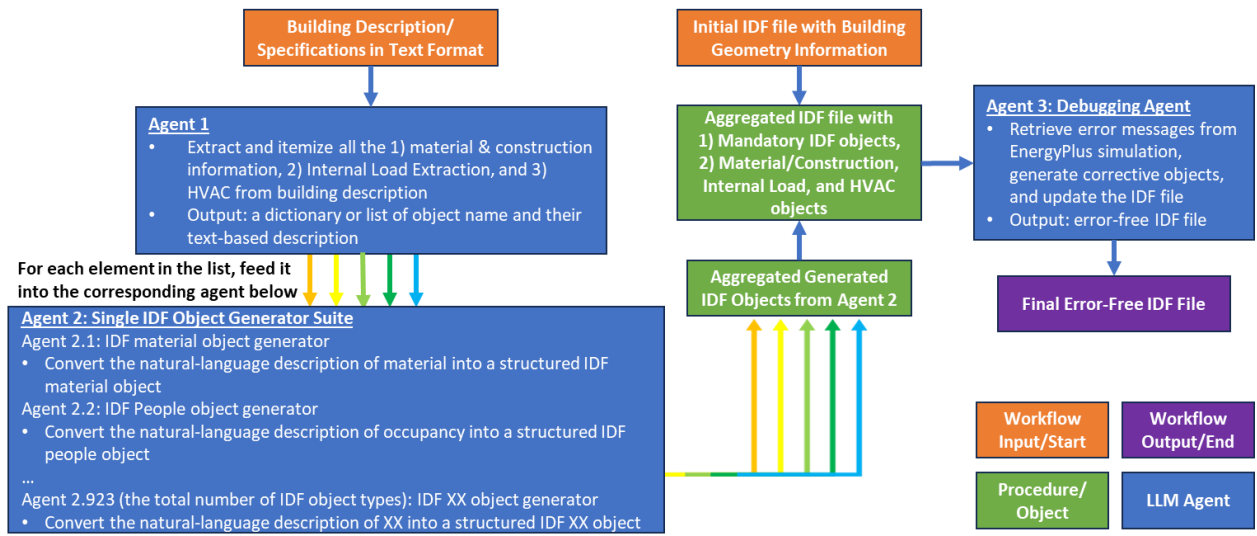

*Figure 2. Diagram of LLM planning workflow to generate IDF file from a building description*

With Agent 1 extracting all objects and their natural-language descriptions in an organized and accurate manner, the next step is to generate individual IDF objects using the Single Object Generation prompt template, which is detailed in Section 3.1.1. This step is handled by Agent 2, which, although collectively referred to as a single entity, consists of 923 different sub-agents. Each sub-agent corresponds to one of the IDF object types. As described in Section 3.1.1, this approach systematically structures the conversion process from natural language descriptions to formal IDF objects, ensuring that all required and optional fields are appropriately addressed and formatted in each IDF object. The process is further refined by Agent 3, the Debugging Agent, which aggregates these IDF objects combined with the initial IDF file with building geometry information and some other mandatory IDF objects pre-created such as "Version", "SizingPeriod:DesignDay", and "RunPeriod" into a single IDF file. Agent 3 runs the aggregated file in EnergyPlus, identifying and correcting errors, primarily those arising from incorrect references. Agent 3 resolves these by adjusting references to ensure consistency across the entire file. The final output of this workflow is the aggregated and debugged IDF file.

In this case study, this workflow was applied to an EnergyPlus modeling task for iUnit, a modular studio apartment experimental test facility located at the National Renewable Energy Laboratory (Figure 3). In terms of LLMs, we utilize GPT model via API called GPT-4 model gpt-4-0613. GPT-4-0613 is an advanced iteration of OpenAI's Generative Pre-trained Transformer models, captured on June 13th, 2023, and optimized for understanding and generating human-like text across a wide range of topics and contexts, with enhanced support for function calling.

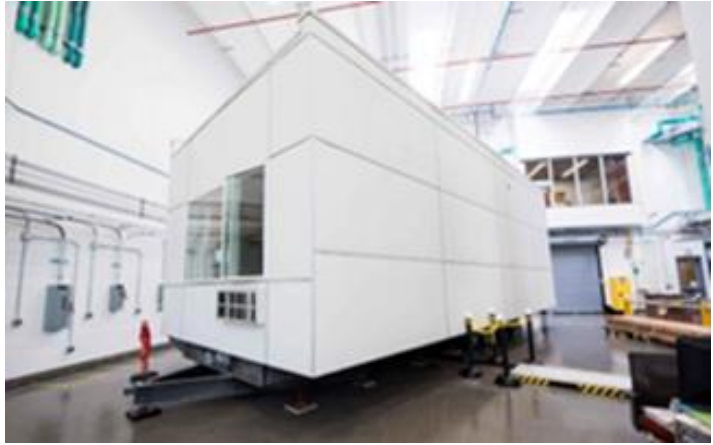
*Figure 3. iUnit floor plan and appearance*

```
Project 1: EnergyPlus Modeling of iUnit
1. Project Description
The goal of Project 1 is to create EnergyPlus Model of iUnit based on the information and measurement
given in this document. iUnit is a modular apartment located at the campus of National Renewable
Energy Laboratory (NREL) that can be moved indoor for guarded testing or outdoor for environmental
testing. NREL is using iUnit to conduct validation-grade energy simulation experiments.

2. iUnit Details
2.3. Constructions
2.3.1. Floor
Table 1. Details on floor construction layers
  Floor Layer 1 (Outside): Rodent Barrier (HDPE)
   Roughness: Smooth; Thickness: 0.05 in; Conductivity: 3.472 Btu·in/hr·ft2·R; Density: 59.3 lb/ft3;
Specific Heat: 0.4347 Btu/lb·R
  Floor Layer 2: Insulated Joist Layer
   Roughness: Smooth: Thickness: 14.462 in; Conductivity: 0.3323 Btu·in/hr·ft2·R; Density: 6.336
lb/ft3; Specific Heat: 0.2357 Btu/lb·R
  Floor Layer 3: OSB
   Roughness: Smooth: Thickness: 1.75 in; Conductivity: 0.826 Btu·in/hr·ft2·R; Density: 33.96
lb/ft3; Specific Heat: 0.289 Btu/lb·R
  Floor Layer 4 (Inside): Plywood
   Roughness: Smooth; Thickness: 0.25 in; Conductivity: 0.8 Btu·in/hr·ft2·R; Density: 33.96 lb/ft3;
Specific Heat: 0.289 Btu/lb·R

2.3.2. Roof
Table 2. Details on roof construction layers
  Roof Layer 1 (Outside): EPDM Rubber
   Roughness: Smooth; Thickness: 0.1 in; Conductivity: 1.3889 Btu·in/hr·ft2·R; Density: 63 lb/ft3;
Specific Heat: 0.48 Btu/lb·R; Solar Absorptance:0.3
  Roof Layer 2: OSB
   Roughness: Smooth; Thickness 0.5 in; Conductivity: 0.8264 Btu·in/hr·ft2·R; Density: 33.96 lb/ft3;
Specific Heat: 0.289 Btu/lb·R
  Roof Layer 3: Insulted Truss Layer
    Roughness: Smooth; Thickness 37.375 in; Conductivity: 0.3425 Btu·in/hr·ft2·R; Density 3.296
lb/ft3; Specific Heat: 0.1900 Btu/lb·R
  Roof Layer 4 (Inside): Drywall
    Roughness: MediumSmooth; Thickness: 0.625 in; Conductivity: 1.1111 Btu·in/hr·ft2·R; Density:
49.94 lb/ft3; Specific Heat: 0.26034 Btu/lb·R

2.3.3. Exterior Wall
Table 3. Details on Exterior wall construction layers
  Wall Layer 1 (Outside): Exterior Finish OSB
   Roughness: Smooth; Thickness: 0.625 in; Conductivity: 0.8264 Btu·in/hr·ft2·R; Density: 33.96
lb/ft3; Specific Heat: 0.289 Btu/lb·R; Solar Absorptance: 0.5
  Wall Layer 2: Rigid Insulation
   Roughness: Smooth; Thickness: 1 in; Conductivity: 0.2 Btu·in/hr·ft2·R; Density: 2.2786 lb/ft3;
Specific Heat: 0.358 Btu/lb·R
  Wall Layer 3: OSB
```

```
  Roughness: Smooth; Thickness 0.5 in; Conductivity: 0.8264 Btu·in/hr·ft2·R; Density: 33.96 lb/ft3;
Specific Heat: 0.289 Btu/lb·R
  Wall Layer 4: Stud Layer
  Roughness: Smooth; Thickness 5.5 in; Conductivity: 0.2927 Btu·in/hr·ft2·R; Density 7.541 lb/ft3;
Specific Heat 0.2573 Btu/lb·R
  Wall Layer 5 (Inside): Drywall
   Roughness: MediumSmooth; Thickness: 0.625 in; Conductivity: 1.111 Btu·in/hr·ft2·R; Density:
49.94 lb/ft3; Specific Heat: 0.26034 Btu/lb·R

2.3.4. Interior Wall
  Interior Wall Layer 1 (Inside): Plywood_2
   Roughness: Smooth; Thickness: 1 in; Conductivity: 0.8 Btu·in/hr·ft2·R; Density: 33.96 lb/ft3;
Specific Heat: 0.289 Btu/lb·R

2.3.5. Window
U-factor = 0.223 Btu/ft2·hr·R
Solar Heat Gain Coefficient = 0.43
Visible Transmittance = 0.42

2.5.6. Door
· Insulated steel exterior door
· Conductivity = 0.20408 BTU in/ft2 F h
· Density of Expanded Polystyrene: 24.0 kg/m^3
· Specific heat of Expanded Polystyrene: 1210.0 J/kg*K

2.4. Internal Loads
2.4.1. People
Internal load "People" strictly follows the National Renewable Energy Laboratory's business hours,
1) Monday-Thursday, 8 a.m.-5 p.m., 2) Friday, 8 a.m.-4 p.m., 3) Closed on Saturday and Sunday.
There are on average two persons in iUnit during business hours and no people in time of the rest.

2.4.2. Lighting
Assume lighting capacity by yourself since it is not measured. This kind of assumption of missing
values is very common in EnergyPlus. Refer to ASHRAE standard. The lighting on/off schedule is
assumed to be the same as the business hours.

2.4.3. Electric Equipment
Assume the capacity by density.

2.5 Infiltration
The infiltration will be defined using IDF object ZoneInfiltration:DesignFlowRate. In this object,
according to the onsite measurements, I = 1, Fschedule = always on (1,1,1,1,…), A = 0.03, B =
0.003, C = 0, D = 0. Apply the infiltration to all zones.

2.6 HVAC System
Use the following HVAC template to design your own HVAC system for iUnit.
- Packaged terminal air conditioner (PTAC) systems with optional hot water boiler
The setpoints of each zone are listed as follows:
- Zone1_Livingroom is conditioned with dual cooling and heating setpoint. The cooling setpoint is
76F and the heating setpoint is 72F.
- Zone2_Bathroom is conditioned with a single heating setpoint. The heating setpoint is 74F.
- Zone3_Storage is an unconditioned room without any setpoints.
```

Agent 1 processes the provided building description, and its output is presented in Table 1. Based on the authors' domain knowledge, the information extracted by Agent 1 is accurate and correctly reflects the building description.

*Table 1. Itemized IDF objects related to materials and construction generated by Agent 2.1 with natural language description from the building description*

| No. | Object Type | Object Name | Natural Language Description |
|-----|-------------|-------------|------------------------------|
| 1 | Material | Rodent_Barrier_HDPE | Roughness: Smooth; Thickness: 1.27 mm; Conductivity: 0.607 W/m·K; Density: 949.3 kg/m3; Specific Heat: 1819 J/kg·K |
| 2 | Material | Insulated_Joist_Layer | Roughness: Smooth: Thickness: 367.3 mm; Conductivity: 0.0461 W/m·K; Density: 101.6 kg/m3; Specific Heat: 986 J/kg·K |
| 3 | Material | OSB | Roughness: Smooth: Thickness: 44.45 mm; Conductivity: 0.1147 W/m·K; Density: 544 kg/m3; Specific Heat: 1209 J/kg·K |
| 4 | Material | Plywood | Roughness: Smooth; Thickness: 6.35 mm; Conductivity: 0.111 W/m·K; Density: 544 kg/m3; Specific Heat: 1209 J/kg·K |

| 5 | Construction | Floor_Construction | Layers: Rodent_Barrier_HDPE, Insulated_Joist_Layer, OSB, Plywood |
|---|---|---|---|
| 6 | Material | EPDM_Rubber | Roughness: Smooth; Thickness: 2.54 mm; Conductivity: 0.243 W/m·K; Density: 1009 kg/m3; Specific Heat: 2008 J/kg·K; Solar Absorptance:0.3 |
| 7 | Material | Insulted_Truss_Layer | Roughness: Smooth; Thickness 949.35 mm; Conductivity: 0.0476 W/m·K; Density 52.8 kg/m3; Specific Heat: 794 J/kg·K |
| 8 | Material | Drywall | Roughness: MediumSmooth; Thickness: 15.875 mm; Conductivity: 0.194 W/m·K; Density: 800 kg/m3; Specific Heat: 1089 J/kg·K |
| 9 | Construction | Roof_Construction | Layers: EPDM_Rubber, OSB, Insulted_Truss_Layer, Drywall |
| 10 | Material | Exterior_Finish_OSB | Roughness: Smooth; Thickness: 15.875 mm; Conductivity: 0.1148 W/m·K; Density: 544 kg/m3; Specific Heat: 1209 J/kg·K; Solar Absorptance: 0.5 |
| 11 | Material | Rigid_Insulation | Roughness: Smooth; Thickness: 25.4 mm; Conductivity: 0.0349 W/m·K; Density: 36.5 kg/m3; Specific Heat: 1497 J/kg·K |
| 12 | Material | Stud_Layer | Roughness: Smooth; Thickness 139.7 mm; Conductivity: 0.0407 W/m·K; Density: 120.6 kg/m3; Specific Heat: 1076 J/kg·K |
| 13 | Construction | Exterior_Wall_Construction | Layers: Exterior_Finish_OSB, Rigid_Insulation, OSB, Stud_Layer, Drywall |
| 14 | Material | Plywood_2 | Roughness: Smooth; Thickness: 25.4 mm; Conductivity: 0.111 W/m·K; Density: 544 kg/m3; Specific Heat: 1209 J/kg·K |
| 15 | Construction | Interior_Wall_Construction | Layers: Plywood_2 |
| 16 | WindowMaterial:SimpleGlazingSystem | Window_Material | U-factor = 1.26 W/m2·K; Solar Heat Gain Coefficient = 0.43; Visible Transmittance = 0.42 |
| 17 | Construction | window_construction_1 | Layers: Window_Material |
| 18 | Material | Insulated_steel_door | Conductivity = 0.356 W/m·K; Density of Expanded Polystyrene: 24.0 kg/m^3; Specific heat of Expanded Polystyrene: 1210.0 J/kg*K |
| 19 | Construction | Door_Construction | Layers: Insulated_steel_door |
| 20 | Schedule:Compact | Business_Hours | Schedule for business hours: Monday-Thursday 8 a.m.-5 p.m., Friday 8 a.m.-4 p.m., Closed on Saturday and Sunday. Use 'ScheduleLimit' as the schedule limit |
| 21 | Schedule:Compact | Always_On | Schedule that is always on (1,1,1,1,...). Use 'ScheduleLimit' as the schedule limit |
| 22 | People | People_iUnit | Average of two persons during business hours, no people at other times. Apply to all zones: ['Zone1_Livingroom', 'Zone2_Bathroom', 'Zone3_Storage']. Schedule: 'Business_Hours'. Activity Level Schedule Name should use 'Always_On' |
| 23 | Lights | Lights_iUnit | Lighting capacity assumed based on ASHRAE standard. Apply to all zones: ['Zone1_Livingroom', 'Zone2_Bathroom', 'Zone3_Storage']. Schedule: 'Business_Hours' |
| 24 | ElectricEquipment | Equip_iUnit | Electric equipment capacity assumed by density. Apply to all zones: ['Zone1_Livingroom', 'Zone2_Bathroom', 'Zone3_Storage']. Schedule: 'Always_On' |
| 25 | ZoneInfiltration:DesignFlowRate | Infiltration_iUnit | Defined using the formula I = 1, Fschedule = always on (1,1,1,1,…), A = 0.03, B = 0.003, C = 0, D = 0. Apply to all zones: ['Zone1_Livingroom', 'Zone2_Bathroom', 'Zone3_Storage']. Schedule: 'Always_On' |
| 26 | HVACTemplate:Thermostat | Thermostat_Livingroom | ThermostatDual setpoint thermostat for Zone1_Livingroom with cooling setpoint at 24.4°C and heating setpoint at 22.2°C |
| 27 | HVACTemplate:Thermostat | Thermostat_Bathroom | Single setpoint thermostat for Zone2_Bathroom with a heating setpoint at 23.3°C |
| 28 | HVACTemplate:Zone:PTAC | PTAC_Livingroom | Packaged Terminal Air Conditioner system for Zone1_Livingroom with dual setpoint control, cooling setpoint at 24.4°C and heating setpoint at 22.2°C |
| 29 | HVACTemplate:Zone:PTAC | PTAC_Bathroom | Packaged Terminal Air Conditioner system for Zone2_Bathroom with a single heating setpoint at 23.3°C |

Each of the 29 elements in the table is then processed by Agent 2, as detailed in Section 3.1.1, resulting in the generation of 29 IDF objects. Due to space constraints and the lengthy nature of IDF objects, we have not included all objects here, but we have attached three below.

```
People,
  People_iUnit, !- Name
  Zone1_Livingroom, !- Zone or ZoneList or Space or SpaceList Name
  Business_Hours, !- Number of People Schedule Name
  People, !- Number of People Calculation Method
  2, !- Number of People
  , !- People per Floor Area
  , !- Floor Area per Person
  0.3, !- Fraction Radiant
  , !- Sensible Heat Fraction
  Always_On, !- Activity Level Schedule Name
  3.82E-8, !- Carbon Dioxide Generation Rate
  No, !- Enable ASHRAE 55 Comfort Warnings
```

```
    ZoneAveraged, !- Mean Radiant Temperature Calculation Type
    , !- Surface Name/Angle Factor List Name
    , !- Work Efficiency Schedule Name
    ClothingInsulationSchedule, !- Clothing Insulation Calculation Method
    , !- Clothing Insulation Calculation Method Schedule Name
    , !- Clothing Insulation Schedule Name
    , !- Air Velocity Schedule Name
    , !- Thermal Comfort Model 1 Type
    , !- Thermal Comfort Model 2 Type
    , !- Thermal Comfort Model 3 Type
    , !- Thermal Comfort Model 4 Type
    , !- Thermal Comfort Model 5 Type
    , !- Thermal Comfort Model 6 Type
    , !- Thermal Comfort Model 7 Type
    , !- Ankle Level Air Velocity Schedule Name
    15.56, !- Cold Stress Temperature Threshold
    30; !- Heat Stress Temperature Threshold

Material,
    Plywood, !- Name
    Smooth, !- Roughness
    0.00635, !- Thickness {m}
    0.111, !- Conductivity {W/m-K}
    544, !- Density {kg/m3}
    1209, !- Specific Heat {J/kg-K}
    0.9, !- Thermal Absorptance
    , !- Solar Absorptance
    ; !- Visible Absorptance

HVACTemplate:Zone:PTAC,
    Zone1_Livingroom, !- Zone Name
    Thermostat_Livingroom, !- Template Thermostat Name
    autosize, !- Cooling Supply Air Flow Rate
    autosize, !- Heating Supply Air Flow Rate
    , !- No Load Supply Air Flow Rate
    , !- Zone Heating Sizing Factor
    , !- Zone Cooling Sizing Factor
    Flow/Person, !- Outdoor Air Method
    0.00944, !- Outdoor Air Flow Rate per Person
    0.0, !- Outdoor Air Flow Rate per Zone Floor Area
    0.0, !- Outdoor Air Flow Rate per Zone
    , !- System Availability Schedule Name
    , !- Supply Fan Operating Mode Schedule Name
    DrawThrough, !- Supply Fan Placement
    0.7, !- Supply Fan Total Efficiency
    75, !- Supply Fan Delta Pressure
    0.9, !- Supply Fan Motor Efficiency
    SingleSpeedDX, !- Cooling Coil Type
    , !- Cooling Coil Availability Schedule Name
    autosize, !- Cooling Coil Gross Rated Total Capacity
    autosize, !- Cooling Coil Gross Rated Sensible Heat Ratio
    3.0, !- Cooling Coil Gross Rated Cooling COP
    Electric, !- Heating Coil Type
    , !- Heating Coil Availability Schedule Name
    autosize, !- Heating Coil Capacity
    0.8, !- Gas Heating Coil Efficiency
    0.0, !- Gas Heating Coil Parasitic Electric Load
    , !- Dedicated Outdoor Air System Name
    SupplyAirTemperature, !- Zone Cooling Design Supply Air Temperature Input Method
    14.0, !- Zone Cooling Design Supply Air Temperature
    11.11, !- Zone Cooling Design Supply Air Temperature Difference
    SupplyAirTemperature, !- Zone Heating Design Supply Air Temperature Input Method
    50.0, !- Zone Heating Design Supply Air Temperature
    30.0, !- Zone Heating Design Supply Air Temperature Difference
    , !- Design Specification Outdoor Air Object Name
    , !- Design Specification Zone Air Distribution Object Name
    None, !- Baseboard Heating Type
    , !- Baseboard Heating Availability Schedule Name
```

```
    autosize, !- Baseboard Heating Capacity
    None; !- Capacity Control Method
```

Since the objects generated from Agent 2 sometimes include placeholders (e.g., TBD1234567890 shown in Section 3.1.1), for example, the placeholder for setpoints schedules in HVAC template, the Debugging Agent (Agent 3) shown below aims to replace these placeholders with the correct reference information for each placeholder. Agent 3 analyzes the .err file from the EnergyPlus simulation, specifically looking for the placeholder "{severe_n_fatal_error_str}" to identify errors related to specific classes and their object names. Using this information, along with the current IDF file referenced by "{current_IDF_file_str}," Agent 3 generates corrected versions of the problematic IDF objects in string format. This ensures that all references are accurately aligned, enhancing the reliability and functionality of the final building energy model.

```
Agent_3 = f""" The following is the fatal or severe error message: ```{severe_n_fatal_error_str}```

Based on the current IDF file (attached in the end), create the corrected IDF objects only for the
wrong objects in string format, and put it between ``` ```, do not include any other things or
comments.

For each IDF object you modify, record its original information in a list of tuples
[('object_type_1', 'object_name1'), ('object_type_2', 'object_name2')], and put it between
///[]///, do not include any other things or comments.

- Specific requirements:
1. The maximum number of IDF objects generated is 10, and the maximum length of the list should
also be less than 10.
2. You must ensure that the other fault-free objects are strictly kept unchanged.

```{current_IDF_file_str}```
    """
```

The final output is the debugged and aggregated IDF objects in one IDF file. We conducted 10 tests of the entire developed workflow to evaluate its success rate in generating accurate and error-free IDF files. All 10 trials successfully produced correct IDF files, achieving a 100% success rate. Additionally, we assessed the time efficiency of the LLM agentic workflow compared to two human modelers: a student new to BEM (who took a BEM class and modeled the iUnit as his final project) and an experienced modeler who built the official iUnit model. The student required two weeks to construct and debug the model, with debugging taking up more than two-thirds of this time. The experienced modeler completed the model in one day. In contrast, the workflow completed the model in just 6 minutes.

### 3.2 Simulation Output Visualization

The case study explores the use of the code generation ability of LLM in postprocessing and visualizing the simulation output of EnergyPlus. The object virtual building in the case study is a reference building: Large Office developed by the National Renewable Energy Laboratory. More details of the building can be found in [20]. The simulation weather file is "2B_USA_AZ_PHOENIX.epw." We use the model ChatGPT-4 (July 6, 2023, Version) Code Interpreter in this case study.

The prompt is shown below, which first provides a background about the source of the CSV file; then, two visualization tasks are specified.

```
I want to conduct a python-based data visualization based on the CSV file I uploaded. The file is
the simulation output of EnergyPlus in the format of CSV.
First, I want to generate a timeseries stack area plot for all enduses. Decide the color of each
stack by your understanding of the enduses. For example, red for heating, and blue for cooling.
Also, just generate one week of results in January. Add the correct title, axis label, and unit to
make the plot look professional.
Second, I want to generate a plot with subplots, each subplot is timeseries line plot of all the
weather variables. Just generate two weeks of results in August. Add the correct title, axis label,
and unit to make the plot look professional.
```

Figure 4 shows the first plot LLM generated, which perfectly matches the description of the stacked plot for enduses in every aspect.

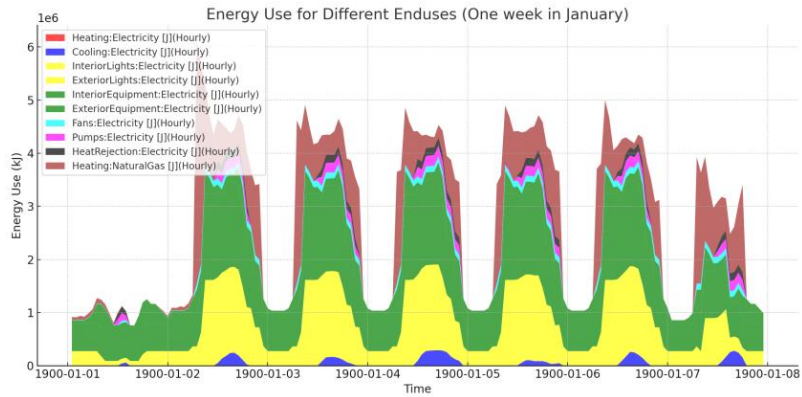

*Figure 4. LLM generated stack area plot of enduses*

Figure 5 shows the LLM-generated subplots of weather conditions, which is the second visualization request in the prompt. The lines in the subplots are created correctly, but the y labels in those subplots overlap with each other. As a result, a follow-up prompt "the y labels in those subplots just overlapped with each other. Please refine the plot" is further sent to LLM and Figure 6 is the improved version, which LLM explains that it "1) rotates the y-labels by 45 degrees, 2) adds more spacing between subplots, and 3) reduces the font size of the y-labels", which successfully address the overlapping problem.

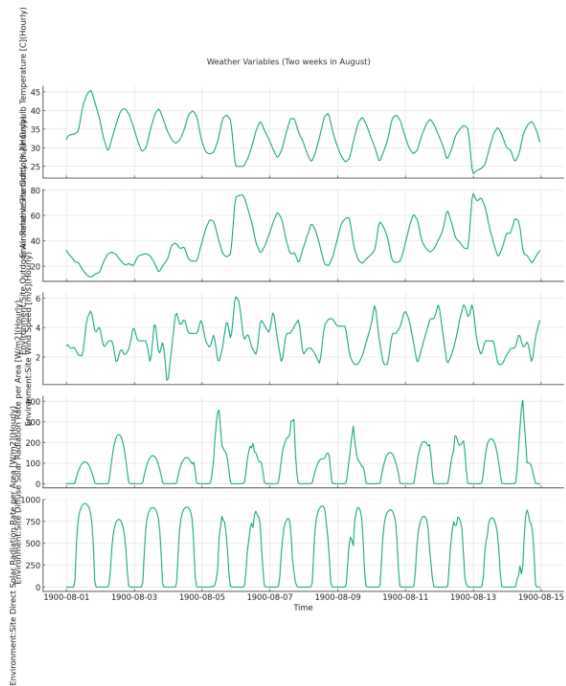

*Figure 5. LLM generated subplots of weather conditions: initial version*

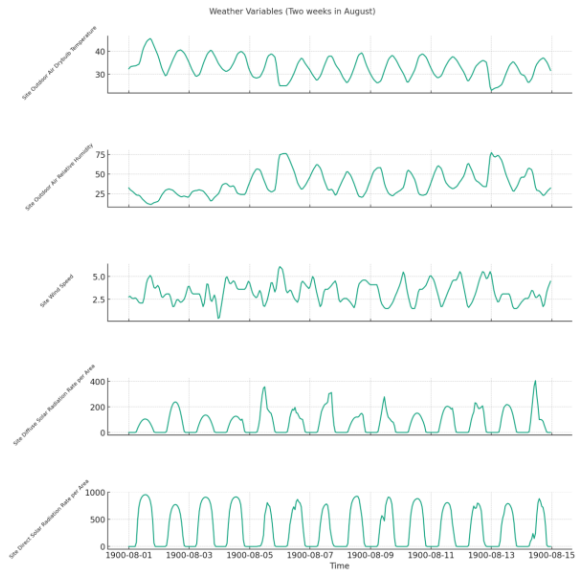

*Figure 6. LLM generated subplots of weather conditions: improved version*

LLM exhibits remarkable accuracy and robustness in visualization when the users clearly understand and describe their visualization objectives. Besides, the human-in-the-loop validation process can also quickly address shortcomings and improve visualization quality.

The time taken to process data and generate plots, which is a key index to evaluate the LLM performance, is significantly reduced from over 15 minutes (based on authors' Python skills) to manually create a Python code to prepare the data for plotting and set the figure configurations, to less than 1 minute to create the prompt by describing the need. Meanwhile, the high accuracy of code generation as well as the fast feedback-based error-fixing feature of LLM make the quality of visualization the same and even better than manual processing.

### 3.3 Simulation Knowledge Extraction and Training

In this case study, we utilize LLM to transform existing knowledge bases into interactive learning platforms, thereby optimizing the educational experience for BEM learners of all proficiency levels by intelligently generating, reorganizing, and presenting information in a user-friendly manner. Specifically, we use retrieval-augmented generation, or RAG, which leverages the synergy of retrieval mechanisms and generative models to dynamically fetch and integrate external knowledge into the response generation process [21]. We use RAG to make BEMcyclopedia (https://bemcyclopedia.com/), a U.S. Department of Energy-sponsored BEM information and education portal, more interactive with BEM learners and users.

The diagram of RAG is shown in Figure 7. In the Indexing Pipeline, we prepare the document by crawling texts from https://bemcyclopedia.com/ and chunking the document into pieces (paragraphs) and generates the embedding (vector representation) for these chunks. The embeddings are then added to a vector database. Indexing Pipeline is a one-time offline process. Next in the Search Pipeline, two steps are involved. Step 1: Vectorize the search query (question) and match it with the most similar text chunks from the Vector Database. Step 2: The LLM is fed with the question and the matching text chunks as context for it to process. This way the LLM only has to most relevant chunks and returns the results in a timely fashion. Since the Search Pipeline is an online process, the two steps are executed each time the user asks a question.

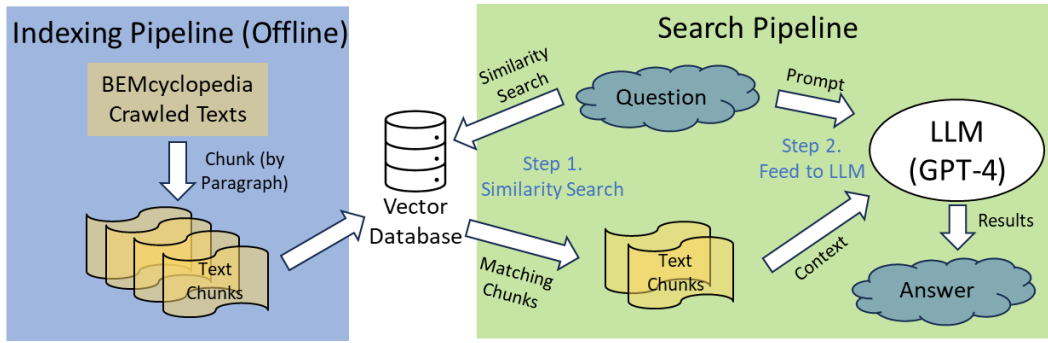

*Figure 7. Diagram of RAG to interact with BEMcyclopedia*

To validate the effectiveness, we compare 1) RAG, 2) ChatGPT (September 25, 2023, Version), and 3) manual search in terms of their ability to answer the questions related to the content in BEMcyclopedia. **The question** we designed is *"What is the purpose of BEMcyclopedia,"* and the **answer from RAG** is shown as follows.

> *The purpose of BEMcyclopedia is to serve as a free resource for energy modelers to learn about BEM and share knowledge with others. According to the website's About page, the online, open-content platform intends to provide practical guidance to practitioners on how, when, and why to apply BEM at various stages of the design process and how to use such analyses to inform design decisions. The content is organized into two main sections: (1) "Learn by Design Task" which is a collection of educational resources organized around a typical building design project's timeline, and (2) "Essential BEM Concepts" which is a collection of general knowledge and fundamental skills applicable to various BEM tasks. The site's scope is focused on building performance analysis, with a specific emphasis on energy and emissions.*

This answer is very comprehensive and includes content across many parts of BEMcyclopedia. When we ask **ChatGPT, the answer** is shown as follows.

> *As of my last update in September 2023, I don't have any information about a "BEMcyclopedia." It's possible that it is a specific resource or tool that has been developed after that date, or it could be a niche or specialized resource that hasn't been widely covered in publicly available data.*

Obviously, the GPT-4 model has not included BEMcyclopedia into its training process, which further showcases the advantage of the RAG feature of including the latest knowledge. In terms of the **manual search**, we use Google (September 2023) as the search engine and "What is the purpose of BEMcyclopedia" as search keywords, and we can find on the webpage that

> *BEMcyclopedia was created as a free resource for energy modelers to learn about BEM, and to share their knowledge with others*

which is far less comprehensive than the answer from RAG.

This method can be further extended to 1) BEM software documentation such as EnergyPlus Input Output Reference, and 2) a combination of multiple BEM training and documentation resources.

## 4 Results and Discussion

### 4.1 Highlighted Observations

This paper explores the potential applications of LLM in BEM. The case studies are designed to provide a preliminary examination of each topic. Based on our findings, several observations are highlighted. **First**, the case studies illustrate the efficacy of incorporating LLMs in various BEM tasks. Visualization of simulation output was particularly successful due to LLM's adept code generation capabilities, simplifying data analysis and plotting with Python code. Knowledge extraction and training from simulations required the use of RAG, adding another layer of complexity. The most intricate task was simulation input generation, which demanded the integration of multiple LLMs with prompt templates to create a multi-agent system for modifying IDF files. Overall, despite their preliminary nature, all case studies were successfully executed and achieved their intended outcomes.

**Second**, the case studies demonstrate that selecting the right LLM techniques is essential to enhance performance and reduce engineering efforts. Besides direct use of LLM, three LLM techniques were utilized: 1) prompt engineering, 2) RAG, and 3) multi-agent LLMs. The complexity and nature of tasks dictate the selection of appropriate LLM techniques. As highlighted in Section 3.3, RAG proved superior to the direct use of LLM. Researchers and engineers should pinpoint the most effective approach among LLM techniques for varied tasks, rather than uniformly applying a single method. In summary, for tasks involving code generation, directly utilizing LLMs typically suffices. When external knowledge is necessary, employing RAG and fine-tuning can effectively handle the task at hand. For more complex such as the case study in Section 3.1.3, multi-step, and hierarchical processes, well-designed prompt engineering and potentially the use of multi-agent LLMs are recommended to navigate the intricacy.

## 4.2 Observed Limitations

While the results are promising, certain limitations are evident. **First**, LLMs require significant computational power, leading to high energy consumption and potential financial burdens, especially when dependent on API-based solutions or necessitating investment in computational hardware like Graphics Processing Units. However, the landscape is changing rapidly, with technological advancements ushering in more efficient and cost-effective LLMs. A prime example is the Llama-2 [22] 7B version, which holds the promise of broadening accessibility and application across diverse fields with low computation cost. For different applications, the impact of computational power varies. For more complex BEM tasks, such as the one introduced in Section 3.1.3, the cost is relatively high, at around $2 per run based on the rate of the GPT-4 model (gpt-4-0613). However, for the other two case studies in Sections 3.1.1 and 3.1.2, the costs are less than $0.1 per run, indicating that simpler tasks are significantly more cost-effective.

**Second**, self-consistency issue, the tendency of the model to provide different or contradictory responses to the same query, was identified as a challenge affecting the reliability and accuracy of the results. In the case studies, we addressed this issue by adjusting the "temperature" parameter of the LLM to zero, although further discussions on alternative solutions were beyond our scope. BEM professionals should remain vigilant and account for these potential inconsistencies when leveraging LLMs in applications requiring high accuracy. Research efforts are actively underway to fundamentally improve the self-consistency of LLMs [23]. These theoretical advancements are crucial in paving the way for more reliable LLMs. However, practical measures are just as vital. Employing continuous validation, corroborating with additional data sources, and seeking expert insights are key strategies to mitigate uncertainties and bolster the reliability of results derived from LLMs.

**Third**, the lack of discussion on fine-tuning is a significant limitation of this study. Fine-tuning is a vital aspect of LLMs, especially for tasks that require in-depth domain knowledge. This process involves refining the model on specialized datasets to enhance its performance. A notable example is the development of BloombergGPT [24], which is a specialized LLM for the financial sector, trained on a diverse range of financial data. The idea of creating a similar model, such as "BEMGPT," is intriguing and holds potential for the field of BEM by eliminating the need for RAG and prompt engineering, which reduces computation cost and engineering effort. However, fine-tuning is even more computationally demanding and poses significant challenges in data design and preparation for training. The creation of a domain-specific model like 'BEMGPT' would require meticulously curated datasets that accurately represent the complexities of BEM. This necessity to refine and adapt LLMs to the specific needs of BEM through fine-tuning presents an important future research direction worth exploring.

**Fourth**, we acknowledge the oversight in addressing the challenges associated with processing long sequences of prompts and managing substantial volumes of formatted text inputs in LLM applications for BEM, especially the application of the simulation input generation. This gap highlights a critical area for future research. To mitigate these issues, future work could explore the implementation of a multi-agent LLM framework. Such a system, featuring a central agent for segmenting extensive text into smaller portions for individual processing and subsequent aggregation, could significantly enhance the handling of large-scale text inputs. Furthermore, RAG can include huge amount of external information outside of LLM prompts, which will not be limited by the input length and token limit. Additionally, the potential of LLMs such as Claude (www.claude.ai), known for their capability to process long inputs, warrants further investigation.

**Fifth,** while LLMs exhibit substantial promise in code generation and improving the usability of software like EnergyPlus, their integration with BEM visual modeling remains an underexplored domain. Current multi-modal LLMs are primarily geared towards creative tasks rather than precise visual modeling, making them less suitable for geometry creation and visualization in BEM applications. This limitation is partly due to their lack of spatial awareness, a critical component for accurate BEM visual modeling. Current technologies do not sufficiently support the spatial reasoning needed for BEM applications. Therefore, although the potential applications of LLMs in visual modeling are exciting, significant advancements are required to harness their full capabilities in this domain. Future research should focus on enhancing the spatial understanding of multi-modal LLMs to improve their accuracy and reliability in visual modeling tasks.

**Sixth**, systematic validation of LLM performance is essential for ensuring robustness and consistency across different users and scenarios. This can involve various methods, such as round-robin exercises, where multiple users independently test the model with different prompts. Such validation techniques are crucial for addressing issues like self-consistency and hallucination, where the model may produce contradictory responses or generate inaccurate information. By employing diverse validation methods, we can comprehensively assess the model's reliability and effectiveness in real-world applications. Given the early exploratory nature of this study, future research should include these systematic validation approaches to thoroughly evaluate LLM performance. This will help identify potential weaknesses, ensure the model's practical applicability in BEM scenarios, and guide improvements for better accuracy and user satisfaction. Implementing continuous validation, corroborating with additional data sources, and seeking expert insights are key strategies to bolster the reliability of results derived from LLMs.

## 5. Conclusion

This paper explores the integration of LLMs in BEM by examining potential applications identified through a literature review of various modeling techniques. The paper highlights the potential of LLMs to address the significant reliance on expert knowledge in BEM, proposing applications including 1) simulation input generation, 2) simulation output analysis, and visualization, 3) conducting error analysis, 4) co-simulation, 5) simulation knowledge extraction and training, and 6) simulation optimization.

**In case studies**, we observed their effectiveness across a range of tasks, from simplifying data analysis with code generation, to integrating multiple LLMs in a multi-agent system for intricate simulation input generation. Crucially, selecting the right technique—be it direct use, prompt engineering, RAG, or multi-agent systems—is paramount to optimize performance and minimize engineering efforts. While LLMs present immense promise, there are challenges, including their significant computational demands and potential self-consistency issues. However, ongoing technological advancements and research efforts are actively addressing these limitations, thereby broadening the scope and ease of LLM applications in diverse fields.

**In the future**, the integration of LLM and BEM will play a crucial role in advancing sustainable and energy-efficient building designs. Collaborative research between AI and building modelers is key to effectively utilizing LLMs in enhancing BEM. This interdisciplinary approach will address the gap between LLM capabilities and the specific needs of BEM, leveraging domain-specific knowledge from model experts alongside the expertise of AI specialists in complex LLM modeling. Although much of AI expertise is currently focused on sectors like medical science and commerce, establishing incentives and raising awareness is necessary to redirect attention and contributions from AI experts to the building sector. The creation of specialized LLMs, such as "BEMGPT," specifically tailored for BEM, holds promise for the future of sustainable building solutions.

## Nomenclature

AI: artificial intelligence

API: Application Programming Interface

BEM: building energy modeling

HVAC: heating, ventilation, and air conditioning

IDD: input data dictionary

IDF: input data file

LLM: large language model